\documentclass[
 reprint,
superscriptaddress,
nofootinbib,
bibnotes,
 amsmath,amssymb,
 aps,
pra,
longbibliography]{revtex4-1}
\usepackage[plainpages = false, pdfpagelabels, 
                 bookmarks,
                 bookmarksopen = true,
                 bookmarksnumbered = true,
                 breaklinks = true,
                 linktocpage,
                 colorlinks = true,
                 linkcolor = blue,
                 urlcolor  = blue,
                 citecolor = blue,
                 anchorcolor = green,
                 hyperindex = true,
                 hyperfigures
                 ]{hyperref} 
\usepackage{verbatim}
\usepackage{enumerate}
\usepackage{makecell, multirow}
\usepackage{tikz}
\usetikzlibrary{patterns}
\usepackage{slashed}
\usepackage{braket}
\usepackage{bbold}
\usepackage{bm}
\usepackage{booktabs}
\usepackage[caption=false,singlelinecheck=off]{subfig}
\usepackage{blindtext}
\usepackage{bbm}
\usepackage{graphicx}
\usepackage{dcolumn}
\usepackage{bm}
\usepackage[mathlines]{lineno}
 \usepackage{esint}
\definecolor{ballblue}{rgb}{0.13, 0.67, 0.8}

\begin{document}


\title{Theoretical proposal to obtain strong Majorana evidence from scanning tunneling spectroscopy of a vortex with a dissipative environment}

\author{Gu Zhang}
\affiliation{Beijing Academy of Quantum Information Sciences, Beijing 100193, China}
\affiliation{State Key Laboratory of Low Dimensional Quantum Physics, Department of Physics, Tsinghua University, Beijing, 100084, China}

\author{Chuang Li}
\affiliation{School of Physics, Huazhong University of Science and Technology, Wuhan, Hubei 430074, China}

\author{Geng Li}
\affiliation{Beijing National Center for Condensed Matter Physics and Institute of Physics, Chinese Academy of
Sciences, Beijing 100190, PR China}

\author{Can-Li Song}
\affiliation{State Key Laboratory of Low Dimensional Quantum Physics, Department of Physics, Tsinghua University, Beijing, 100084, China}
\affiliation{Frontier Science Center for Quantum Information, Beijing 100184, China}

\author{Xin Liu}
\email{phyliuxin@hust.edu.cn}
\affiliation{School of Physics, Huazhong University of Science and Technology, Wuhan, Hubei 430074, China}
\affiliation{Hefei National Laboratory, Hefei 230088, China}

\author{Fu-Chun Zhang}
\email{fuchun@ucas.ac.cn}
\affiliation{Kavli Institute for Theoretical Sciences, University of Chinese Academy of Sciences, Beijing 100190, China}
\affiliation{Hefei National Laboratory, Hefei 230088, China}

\author{Dong E. Liu}
\email{dongeliu@mail.tsinghua.edu.cn}
\affiliation{State Key Laboratory of Low Dimensional Quantum Physics, Department of Physics, Tsinghua University, Beijing, 100084, China}
\affiliation{Frontier Science Center for Quantum Information, Beijing 100184, China}
\affiliation{Beijing Academy of Quantum Information Sciences, Beijing 100193, China}
\affiliation{Hefei National Laboratory, Hefei 230088, China}

\begin{abstract}

It is predicted that a vortex in a topological superconductor contains a Majorana zero mode (MZM). The confirmative Majorana signature, i.e., the $2e^2/h$ quantized conductance, however is easily sabotaged by unavoidable interruptions, e.g. instrument broadening, non-Majorana signal, and extra particle channels. To avoid these interruptions, we propose to obtain strong Majorana evidence following our novel Majorana hunting protocol that relies on dissipation introduced by disorders at e.g., the sample-substrate interface. Within this protocol, we highlight three features, each of which alone can provides a strong evidence to identify MZM. Firstly, dissipation suppresses a finite-energy Caroli-de Gennes-Matricon (CdGM) conductance peak into a valley, while it does not split MZM zero-bias conductance peak (ZBCP). Secondly, we predict a dissipation-dependent scaling feature of the ZBCP. Thirdly, the introduced dissipation manifests the MZM signal by suppressing non-topological CdGM modes. Importantly, the observation of these features does not require a quantized conductance value $2e^2/h$.


\end{abstract}

\date{\today}
\maketitle

\section{Introduction}

A Majorana zero mode (MZM)~\cite{Read-2000-PRB,Kitaev2001} is an exotic quasi-particle state that has equal electron and hole weights.
Importantly, due to its non-abelian feature, MZM has become one important candidate to realize quantum operations and the ensued topologically protected quantum computation~\cite{kitaev2003fault,nayak2008non,sarma2015majorana}. Some pioneering theories have predicted the existence of MZMs in multiple physically realizable systems~\cite{FLiang-2008-PRL,sato2009PRL,SLTD-2010-PRL,LutchynSauSarmaPRL10,OregRefaelOppenPRL10,Beenakker-2011-PRB,Yazdani-2013-PRB}. Experimental efforts had been made to detect MZMs in many platforms, e.g., semiconductor-superconductor nanowire structures~\cite{mourik2012signatures,deng2012anomalous,das2012zero,finck2013anomalous,churchill2013superconductor,chang2015hard,krogstrup2015epitaxy,DengScience16,Nichele-2017-PRL,vaitiekenas2021zero,HaoZhangPRR22,PanCPL22,WangX22}, the vortex states in iron-based superconductors~\cite{DHong-2018-Sci,FDongLai-2018-PRX,Hanaguri-2019-NatM,ZPeng-2019-NatPhys,KLingYuan-2019-NatPhys,WHaiHu-2019-arXiv,DHong-2020-Sci,DHong-2020-NatCom,FDongLai-2021-PRL,KLingyuan-2021-NatCom,LWenyao-2021-arXiv} or topological insulator-superconductor heterostructures~\cite{JJinFeng-2015-PRL,JJinFeng-2016-PRL,LvSciBulletin17,YuanNatPhys19}, magnetic atom chain~\cite{Yazdani-2014-Sci} and topological Josephson junctions~\cite{fornieri2019evidence,ren2019topological,dartiailh2021phase}, etc.
Among all those platforms, hybrid nanowire structures (theoretically proposed by Refs.~\cite{LutchynSauSarmaPRL10,OregRefaelOppenPRL10}) have received a worldwide interest. However, the Majorana hunting in these systems encounters the emergence of false-positive signals arising from trivial Andreev bound states and other disorder effects (see e.g., Refs.~\cite{PientkaDisorder,LiuPotterLawLeePRL12,pikulin2012zero,prada2012transport,TakeiSarmaPRL13,DongPRB18,PanDasSarmaPRR20,PanLiuWimmerSarma21}). Regarding this debate, Ref.~\cite{DongPRL13} proposed a scheme to suppress these false-positive signals by introducing dissipation to the probe (see recent progress in Refs.~\cite{DonghaoPRL22,ShanZhangPRL22,WeakX22} for more detailed theoretical and experimental analysis). 
Recently, Microsoft quantum team proposed another solution and considered a more sophisticated design by combining three different measurements on a single detection device~\cite{MSprotocol2021,aghaee2022inas}, which however sacrifices the quality of the local measurement signals~\cite{aghaee2022inas}. So far, there is no solid evidence to conclude the existence of the elusive MZM.

Zero-bias conductance peak (ZBCP) has also been reported as a Majorana feature in spectroscopic measurements of vortex states via the standard scanning tunneling microscopy (STM) technique
~\cite{JJinFeng-2015-PRL,JJinFeng-2016-PRL,LvSciBulletin17,DHong-2018-Sci,FDongLai-2018-PRX,Hanaguri-2019-NatM,ZPeng-2019-NatPhys,KLingYuan-2019-NatPhys,WHaiHu-2019-arXiv,YuanNatPhys19,DHong-2020-NatCom,FDongLai-2021-PRL,KLingyuan-2021-NatCom,LWenyao-2021-arXiv}.
More inspiringly, a recent progress~\cite{GaoNature22}
shows that topological vortexes can form into an ordered lattice through the introduction of a biaxial charge density wave. This technical advance can potentially reduce the difficulty of Majorana manipulation: the next stage in the realization of topological quantum computation.
However, the spectroscopically measured signal is normally influenced by interruptions (e.g. instrument broadening, potential non-Majorana signal~\cite{PanDasSarmaPRR20}, and possible existence of extra particle channels~\cite{Colbert-QP-2014}), that may sabotage the quantized Majorana conductance $2e^2/h$~\cite{law2009majorana}.
Actually, most vortex-based researches (e.g., Refs.~\cite{JJinFeng-2016-PRL,DHong-2018-Sci,FDongLai-2018-PRX,Hanaguri-2019-NatM}) report ZBCPs in the unit of a.u. (arbitrary unit).
In addition, the obtained conductance is close to the quantized value only under an extremely small tip-sample distance (see e.g., Refs.~\cite{DHong-2020-Sci,FDongLai-2019-CPL}).
Of this situation, however, the tip has a spatially wide contact with the sample, where the validity of the point-contact tunneling model remains an open question.


\begin{figure*}
  \includegraphics[width=1.95\columnwidth]{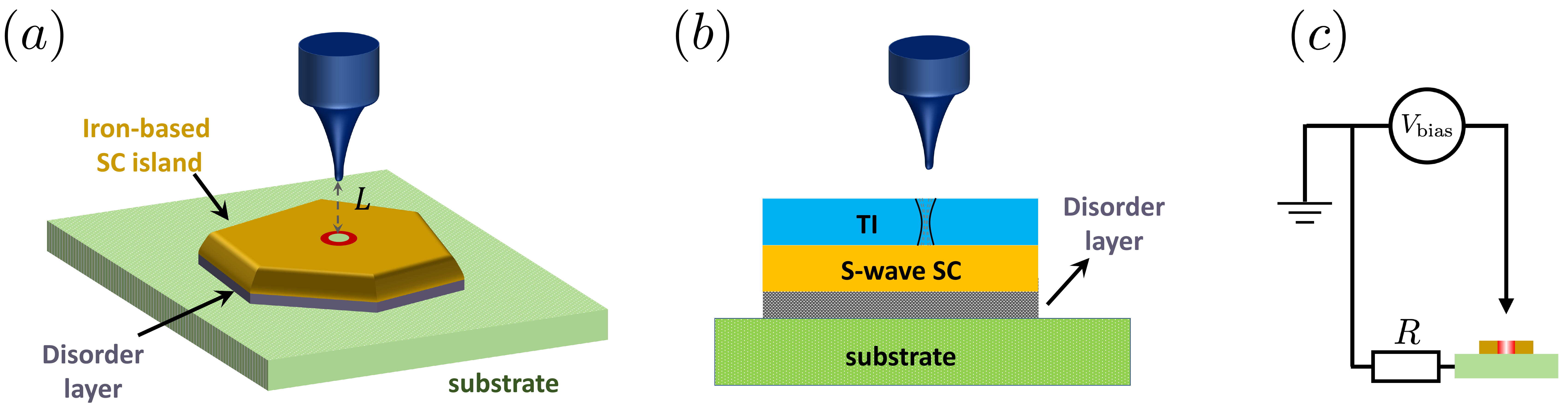}
  \caption{
  Two possible experimental realizations of our theory.
  (a) An iron-based superconducting island placed on top of a substrate. A MZM is expected in the middle of the island (the red circle). Due to the disordered layer, disorders are induced between the island and the substrate, leading to dissipative tip-sample-tunnelings.
  (b) An alternative structure where a topological insulating (TI) layer is placed on top of an s-wave superconductor~\cite{FLiang-2008-PRL,SLTD-2010-PRL}. A disordered layer exists between the s-wave superconductor and the substrate, leading to a dissipative tip-sample tunneling.
  (c) The equivalent circuit of our system. Scatterings at the disordered layer are incorporated by a resistance $R$.
  A voltage bias $V_\text{bias}$ is applied between the tip and the sample.
  }
  \label{fig:sample}
\end{figure*}

In this paper, we solve problems above, by introducing dissipative environments that couple to the electron tunneling between tip and vortex.
Experimentally, dissipation can be introduced via a disorder layer~\cite{BrunPRL12,RoditchevPRL13,JeonKimKukNJP20} or strong lattice mismatch between the superconductor and substrate, illustrated in Fig.~\ref{fig:sample}.
Indeed, a strong dissipation has been observed in the STM detection~\cite{BrunPRL12,RoditchevPRL13,OdobescuPRB15,CarbilletPRB20,JeonKimKukNJP20,RachmilowitzNPJQM20}, including Ref.~\cite{RoditchevPRL13,JeonKimKukNJP20} where dissipation arises from a disordered Pb wetting layer.
Akin to the dissipation influence in a hybrid nanowire system~\cite{DongPRL13,DonghaoPRL22,ShanZhangPRL22,ZhichuanPRB22}, dissipation added to a vortex system suppresses possible interruptions from e.g., soft-gaps or that from a continuous Caroli-de Gennes-Matrico (CdGM) spectrum~\cite{ShorePRL89}.
More recently, Ref.~\cite{OdobeskoPRB20} experimentally demonstrates that the introduction of dissipation (realized via oxidization the Nb layer) can reduce the anisotropy and amplitude of conductance, by suppressing tunnelings into CdGMs.
Although Ref.~\cite{OdobeskoPRB20} mainly focuses on regimes out of the superconducting gap, we anticipate the observation of similar features within the gap, following the analysis of our work.
Remarkably, we also discover strong and vortex-system-specified Majorana evidence for three generic Majorana hunting situations. Importantly, all features predicted in our work do not require the knowledge of the conductance unit or the quantized conductance value~\cite{law2009majorana}, thus greatly reducing the difficulty of possible experiments.
Although disorders at the superconductor-substrate interface is required in our protocol, a clean TI-superconductor interface is required, to support the existence of Majorana zero modes.
In real experiments, a clean TI-superconductor interface (on top of a disordered superconductor-substrate interface) can be possibly realized after optimizing sample fabrications, e.g., increasing the thickness of the superconducting layer.


The observed results in vortex Majorana hunting can be classified into three generic situations: (i) the case with a ZBCP and multiple finite-energy CdGM conductance peaks~\cite{FDongLai-2018-PRX,Hanaguri-2019-NatM,KLingYuan-2019-NatPhys}; (ii) the case with only one ZBCP~\cite{JJinFeng-2015-PRL,LvSciBulletin17,DHong-2018-Sci}, either topological or trivial, and (iii) the case where the Majorana ZBCP is concealed by a continuous CdGM spectrum~\cite{ShorePRL89}.
In the conventional (i.e., dissipation-free) STM spectrum detection, the Majorana signature requires the observation of a robust quantized conductance with a reasonably small tip-sample distance: otherwise it is hard to reach a conclusive statement due to potential interruptions from false-positive signals. As the central point of our work, we show that dissipation provides strong Majorana evidence in a vortex system. 
Of case (i), dissipation strongly suppresses CdGM peaks within the gap, leading to splitting of them.
ZBCP arising from the MZM, by contrast remains.
The distinct responses of CdGM and MZM peaks to dissipation then provide a strong and clear evidence of a Majorana.
Of case (ii), the MZM presence can be proven by measuring the dissipation-dependent universal scaling of the conductance; while the trivial ZBCP splits due to suppression from dissipation.
Of case (iii), conductance from a continuous CdGMs can be suppressed by dissipation, thus manifesting the `dissipation-proof' Majorana signal.



\section{The scheme setup and major results}

We consider the setup of Fig.~\ref{fig:sample} where a sample is spectroscopically detected by a tip.
The sample contains a vortex in the topological regime.
Such a system can be realized with either an iron-based superconductor in the topological regime [Fig.~\ref{fig:sample}(a)]~\cite{DHong-2018-Sci,FDongLai-2018-PRX,Hanaguri-2019-NatM,ZPeng-2019-NatPhys,KLingYuan-2019-NatPhys,WHaiHu-2019-arXiv,DHong-2020-NatCom,LWenyao-2021-arXiv,KLingyuan-2021-NatCom,FDongLai-2021-PRL}, or a topological insulator (TI) on top of a s-wave superconductor [Fig.~\ref{fig:sample}(b)]~\cite{JJinFeng-2015-PRL,JJinFeng-2016-PRL}.
In either case, a disordered layer can exist between the superconducting layer and the substrate.
With the disordered layer, tip-sample tunnelings couple to dissipative modes that can be mimicked by a circuit Ohmic resistance $R$ [Fig.~\ref{fig:sample}(c)].
The dissipative modes, arising from the disordered layer, suppress tip-sample tunnelings. The suppression is especially strong under low energies (i.e., low temperature and bias), when the dissipated energy is unaffordable by the system.
This phenomenon, known as dynamical Coulomb blockade, has been well studied in mesoscopic devices (see, e.g., Ref.~\cite{Ingold1992SUS}), as well as spectroscopically probed superconducting islands~\cite{BrunPRL12,RoditchevPRL13,JeonKimKukNJP20,RachmilowitzNPJQM20}.
For clarification, we emphasize that not all impurities can produce such dissipative modes.
Indeed, impurities at the sample surface are known to generate only trivial localized states that interrupt the Majorana detection. On the other hand, impurities outside of the coherence length, e.g., these in the fridge or the amplifier, can be considered as resistors in-series (see discussions in e.g., Refs.~\cite{ShanZhangPRL22,ZhichuanPRB22}), and are thus irrelevant to our consideration.
In our case, the chosen dissipative resistance $R$ should be large enough to suppress the CdGM peaks (at experimentally accessible temperatures), and also $r \equiv Re^2/h<0.5$ (the dimensionless dissipation) to avoid sabotaging the Majorana ZBCP~\cite{DongPRL13,DonghaoPRL22}.
Dissipation around several kilo-ohms has been experimentally realized by the inter-surface disorders or the generic mismatch between superconductor-substrate lattices~\cite{BrunPRL12,RoditchevPRL13,JeonKimKukNJP20,RachmilowitzNPJQM20,CarbilletPRB20}.
The dissipation strength $r$ in real experiments can be obtained by measuring the scaling feature of the conductance in the normal state.
We also assume a large enough sample such that its charging energy is smaller than other relevant energy scales, including the STM resolution, temperature and bias in the system.



\begin{figure}
  \includegraphics[width=0.8\linewidth]{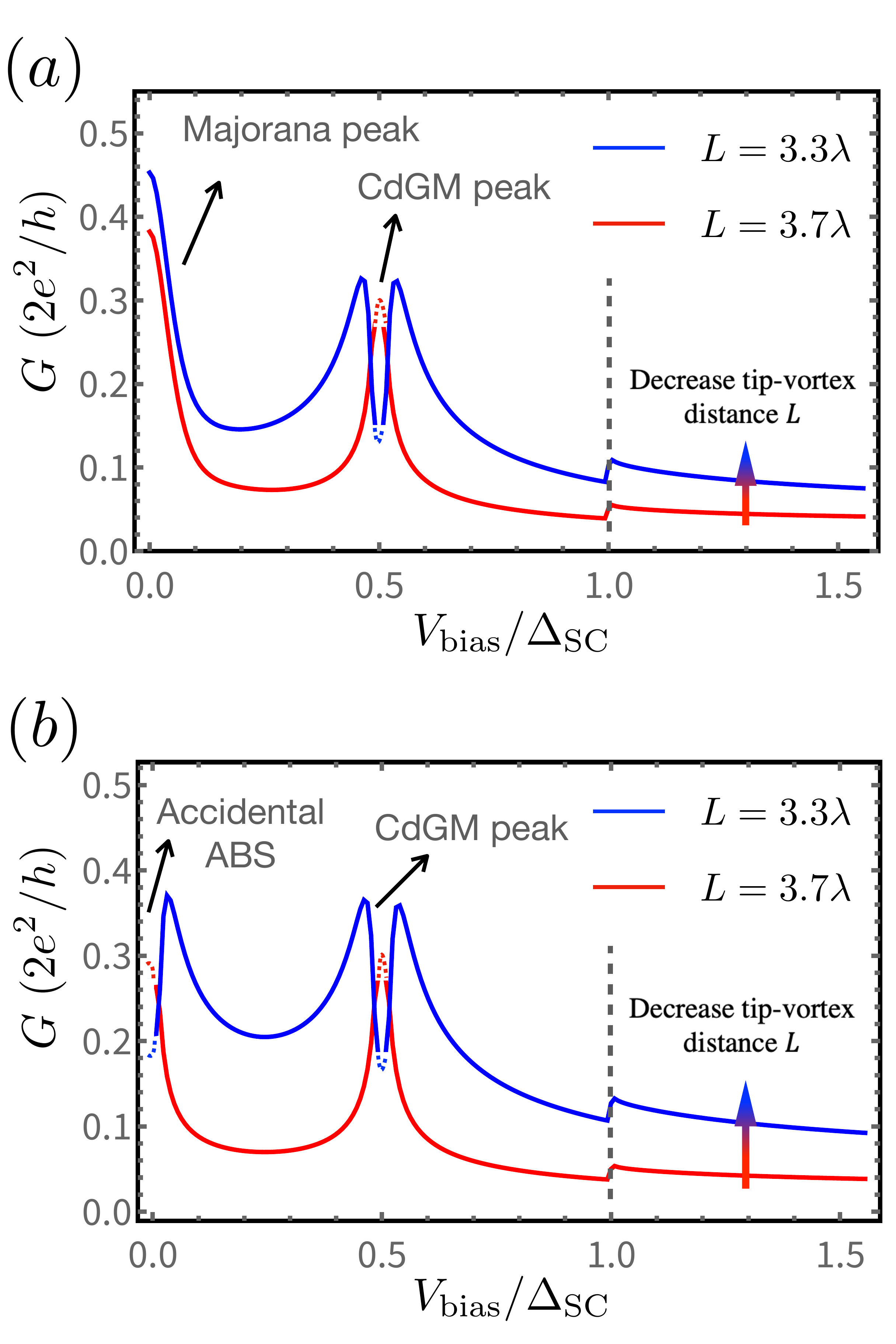}
  \caption{Conductance landscape of dissipative tunneling into a vortex for (a) The topological situation and (b) the trivial situation. In both figures, the red curve refers to the case with a larger tip-sample distance (or a smaller tunneling amplitude), in comparison to that of the blue~\cite{SupMat}.
  Here $\lambda$ refers to the decaying amplitude of the tip-MZM tunneling $t_\text{MZM}(L) \propto \exp(-L/\lambda)$.
  In (a), the zero-bias conductance peak arises from the topological MZM. Its peak value, calculated with Eq.~\eqref{eq:bsg} following the Thermodynamic Bethe Ansatz~(see Ref.~\cite{FendleyPRB95} or a brief introduction in Appendix~\ref{app:tba}), increases monotonously when the tip approaches the sample.
  In contrast, conductance peaks from both finite-energy CdGMs or zero-energy Andreev bound state (denoted as ``Accidental ABS'' in the figure) split for a small enough tip-sample distance (valleys of blue curves).
  The CdGM conductance is evaluated with renormalization group (RG) equations [see Eqs.~\eqref{eq:transmission_flows} and \eqref{eq:rg_final} for the initial and final RG equations, and Appendix~\ref{app:coulomb-gas} for more details]. Close enough to these peaks, the energy from bias difference is smaller than the sample temperature $T = \epsilon_\text{ch}$.
  Conductance in these areas, highlighted by dashed lines, is uncertain from RG method.
  }
  \label{fig:universal_vs_non-universal}
\end{figure}

\begin{figure}
  \includegraphics[width=\linewidth]{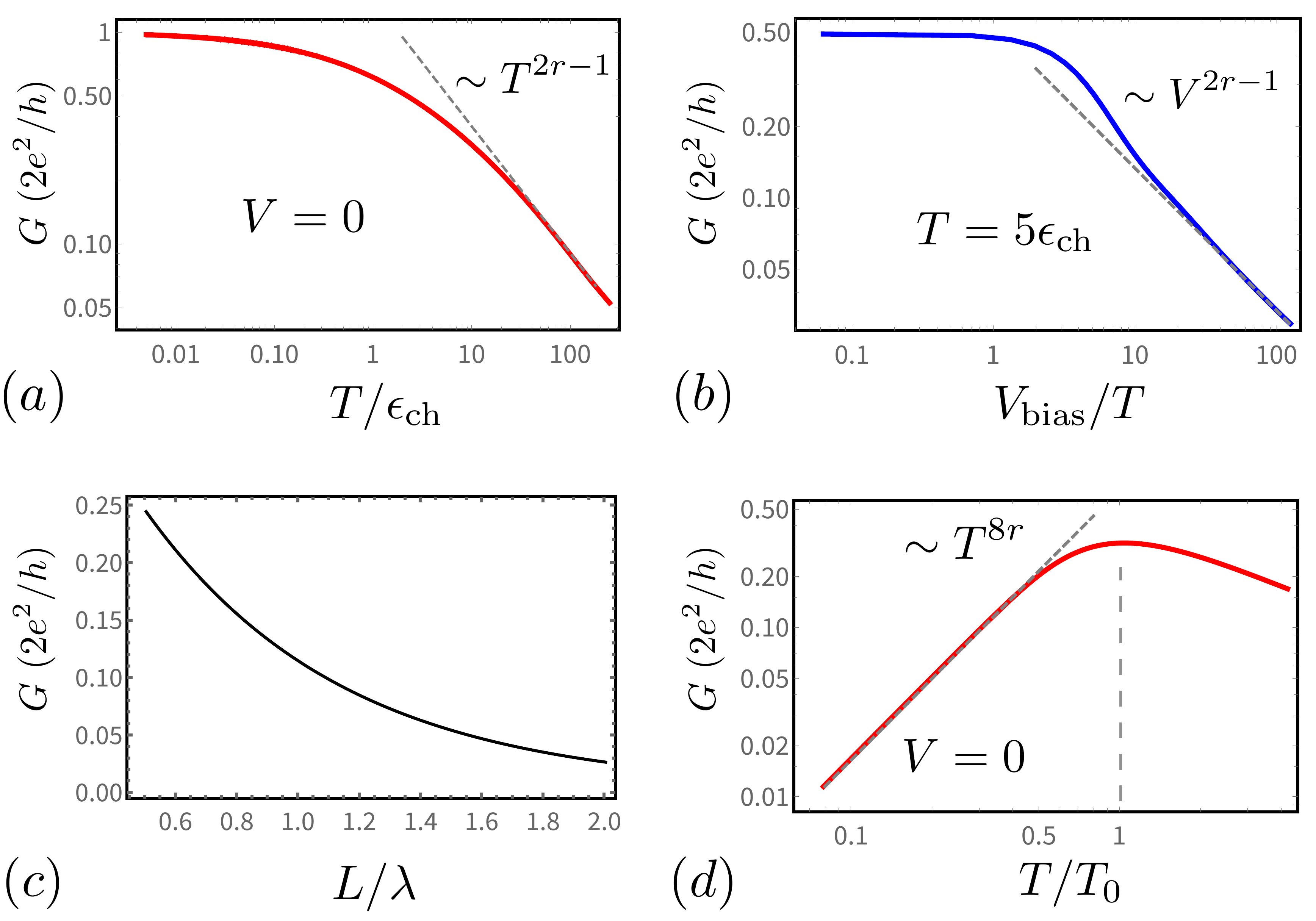}
  \caption{Universal signatures of a  Majorana-induced conductance peak~\cite{SupMat}.
  The universal feature can be observed by changing either the temperature [(a), at zero bias $V = 0$], or the bias [(b), with a fixed temperature $T = 5\epsilon_\text{ch}$]. Of the latter case, the universal feature appears when $V_\text{bias}$ is much larger than the temperature $T =5 \epsilon_\text{ch}$.
  When $V_\text{bias} \ll T $, the conductance saturates to a value smaller than $2e^2/h$, due to the finite-temperature effect.
  The Majorana coupling $t_\text{MZM}$ has the same value for (a) and (b).
  The universality might be alternatively detected by measuring the zero-bias conductance as function of the tip-sample distance $L$ (c), after obtaining the dependence of tunneling amplitude on $L$, i.e., $t_\text{MZM}(L)$.
  In making (c) we have assumed the tip-sample tunneling $t_\text{MZM} \propto \exp(-L/\lambda)$, as an easy example.
  All curves are obtained following Eq.~\eqref{eq:bsg}, following the Thermodynamic Bethe Ansatz~\cite{FendleyPRB95}.
 (d) For an accidental zero-energy Andreev bound state, the equilibrium (i.e., $V = 0$) conductance decreases at a low temperature, following a different scaling feature $G \sim T^{8r}$. Here $T_0$ refers to the temperature with the highest conductance.
  }
  \label{fig:universal_features}
\end{figure}

Generically, a vortex in a topological superconducting island contains a zero-energy Majorana, and multiple finite-energy CdGMs.
These CdGMs can be considered as Andreev bound states that are recognizable in the energy space (see e.g.,~\cite{FDongLai-2018-PRX,Hanaguri-2019-NatM,KLingYuan-2019-NatPhys} for experimental realizations). Fortunately, dissipation influences received by these finite-energy CdGM peaks are distinct from that of the Majorana ZBCP (see Appendixes \ref{app:coulomb-gas}, \ref{app:tba}, and Supplementary Information~\cite{SupMat} for the details on the calculation of CdGM peaks and the Majorana ZBCP, respectively). This fact, as will be shown below (the details of the theoretical analysis and calculation will be shown in the next section), provides us a strong Majorana evidence. We first discuss our improved Majorana detection scheme and main results, and leave the details of the calculation in the next section.
Without the loss of generality, the calculations are carried out with dissipation $r \equiv Re^2/h= 0.2$. We emphasize that features predicted below are also anticipated for other finite dissipation, as long as $r<0.5$.
We define an energy scale $\epsilon_\text{ch} \propto t_\text{MZM}^{2/(1 - 2 r)} \tau_c^{(1 + 2r)/(1 - 2 r)} $ that depends on the tunneling amplitude $t_\text{MZM}$ [of the Hamiltonian~\eqref{eq:ht_mbs}], dissipation $r$ and the inverse of the ultra-violate cutoff $\tau_c$ ($\sim 1/E_F$, with $E_F$ the fermi energy).
The value of $\epsilon_\text{ch}$ is experimentally tunable by changing the tip-sample distance.
Theoretically, $\epsilon_\text{ch}$ labels the energy above which scaling feature begins to develop.
As an experimentally reasonable value, We take the scale $\epsilon_\text{ch} = \Delta_\text{sc}/400$ (in both Figs.~\ref{fig:universal_vs_non-universal} and \ref{fig:universal_features}), where $\Delta_\text{sc}$ is the bulk superconducting gap.
This corresponds to an experimentally accessible sample temperature $T = 5 \epsilon_\text{ch} \approx 500$mK~\cite{JJinFeng-2016-PRL,FDongLai-2018-PRX,Hanaguri-2019-NatM,DHong-2020-NatCom,KLingyuan-2021-NatCom} in real experiments, considering the typical iron-based superconducting gap $\sim 35$K.
If needed, temperatures lower than 100mK are also experimentally realizable~\cite{HanaguriRevSciIns18,Hanaguri-2019-NatM,MartaRevSciIns21}.



More specifically, in Fig.~\ref{fig:universal_vs_non-universal}, we compare the conductance landscape of the topological (a) and trivial (b) situations.
Of the trivial case, the ZBCP is assumed to come from a zero-energy Andreev bound state.
We plot only conductance of positive bias, as the landscape is symmetric in bias.
In both figures, the blue curves are plotted with a smaller tip-sample distance --- equivalently, a stronger tip-sample tunneling than those of the red curves.
In Fig.~\ref{fig:universal_vs_non-universal}(a), the conductance contains contributions from both the zero-energy Majorana and a finite-energy CdGM (centered at $V_\text{bias} = 0.5 \Delta_\text{sc}$).
Noticeably, when the tip-sample distance becomes small, the CdGM conductance peak (the red curve) can split into a valley (the blue curve). This peak-valley transition signifies the influence of dissipation on the tunneling into a CdGM.

Experimentally, one can fix the experimental temperature $T_\text{exp}$, and observe the peak splitting by decreasing the tip-sample distance $L$.
In this work we treat the tip-sample junction as a tunneling barrier, with which the tip-sample transmission $\Gamma_0 (L) \propto \exp(-2L/\lambda)$ decays exponentially when $L$ increases. The decaying width $\lambda$ reflects the feature of the tunneling potential barrier.

As a natural question, what is the requirement on the tip-sample distance $L$, to observe the peak splitting at a given experimental temperature $T_\text{exp}$? To answer this question, we provide the following experimental steps. As the first step, one measures the transmission $\Gamma_0 (L)$ at a temperature $T_0$. This temperature is chosen to be high enough such that (within the range of $L$ during this measurement) interruption from dissipation can be avoided.
Then, we switch back to the experimental temperature $T_\text{exp} < T_0$, where dissipation will play an important role.
At $T_\text{exp}$, we expect to see the CdGM peak-splitting as long as $L<L^{*}$,where the critical distance $L^*$ is defined by its corresponding transmission that satisfies $\Gamma_0 (L^{*}) = \ln[1/(1-\beta)]/\ln (T_0/T_\text{exp})$ (see Appendix~\ref{app:spliting_energy}).
Factor $\beta < 1$ is non-universal and depends on the position detected by the tip: $\beta$ is close to $(1-2r)/(1+2r)$ in case of an extremely large asymmetry among different weights of the CdGM at the tip position.
Here, the weight specifically means the weight between different components of the CdGM wavefunction
[see Eq.~\eqref{eq:ht_abs} and Appendix~\ref{app:coulomb-gas}].
For instance, one can see the peak splitting of CdGM peaks when $L$ changes from $3.7\lambda$ to $3.3\lambda$ (see Fig.~\ref{fig:universal_vs_non-universal}). The curves are evaluated by solving the RG equations numerically.

By contrast, when the total electron and hole weights equal,
$\beta$ approaches one, where the CdGM conductance signature at the detected position strongly mimics that of a real Majorana. Fortunately, the symmetric weight situation for a non-Majorana CdGM is not robust and can be only preserved at some special tip locations. Therefore, a CdGM with accidental symmetric weight can be distinguished (from a Majorana) by detecting its tunneling signal at different positions.

In strong contrast, the zero-energy Majorana peak remains, and keeps growing when decreasing the tip-sample distance.
We emphasize that for a real MZM, this Majorana-specified feature does not depend on the tip position (in the 2D plane parallel to the sample). Indeed, the dissipative tunneling into a MZM is anticipated as a universal process~\cite{DongPRL13,DonghaoPRL22}, and increases monotonously with an increasing tip-sample tunneling.
The coexistence of a Majorana ZBCP and splitted CdGM peaks is a strong Majorana signature [Fig.~\ref{fig:universal_vs_non-universal}(a)].
In strong contrast, if the ZBCP arises instead from an accidentally zero-energy Andreev bound state, the ZBCP always splits if one reduces the tip-sample distance [Fig.~\ref{fig:universal_vs_non-universal}(b)].
We emphasize that dissipation is crucial in this landscape measurement --- otherwise the CdGM and Andreev bound state peaks will not split.
The observation of the proposed phenomenon above, importantly, does not require knowing the conductance unit. Calibration is thus unnecessary, lowering the difficulty of its potential experimental realization.
A concrete observation of the peak splitting requires a STM resolution to be smaller than the splitting temperature.
This temperature describes an energy scale at which the dissipation effect starts to cause the peak splitting as the temperature drops further.
To date, STM resolution around or smaller than $100$mK has been realized~\cite{HanaguriRevSciIns18,MartaRevSciIns21}.
Given large enough dissipation, this resolution is smaller than the peak-splitting temperature [for instance, see the evaluated peak-splitting temperature ($\sim 1$K) in the previous paragraph], thus being
enough to observe the peak splitting.

\begin{figure*}
  \includegraphics[width=2\columnwidth]{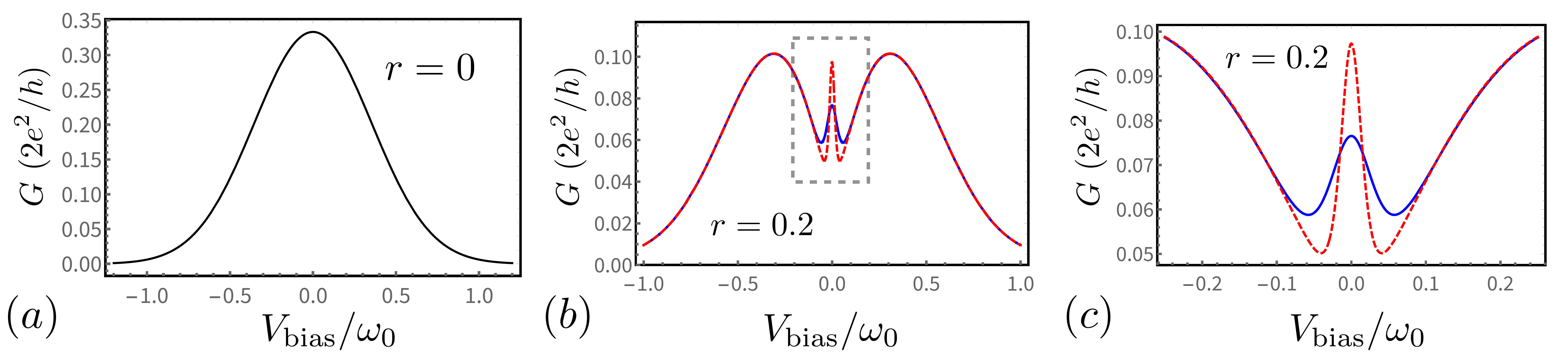}
  \caption{The conductance of a vortex that contains both a zero-energy Majorana and a continuous spectrum of CdGM~\cite{SupMat}. Energy scale $\omega_0$ is the half-width of the continuous spectrum. (a) The $r = 0$ case. The Majorana signal is disguised by the CdGM conductance background. Here $\omega_0$ refers to the range of energy of CdGM spectrum. (b) The $r = 0.2$ case. The Majorana ZBCP becomes visible after the CdGM conductance has been suppressed by dissipation. The red and blue curves correspond to conductance curves with a comparatively lower and higher temperature, respectively. (c) The zoom-in plot of the conductance curves in the grey dashed box. When temperature decreases, the peak-valley difference of the ZBCP increases, leading to a more manifest Majorana signal. Here we calculate the conductance from the continuous spectrum following the $P(E)$ theory Eq.~\eqref{eq:pe}.
  We emphasize that the ZBCP, while being small, appears under the influence of dissipation. It is intrinsically different, although appearing similar, in comparison to the pioneering Majorana hunting achievement in a hybrid nanowire system~\cite{mourik2012signatures}. Indeed, with dissipation present, possible interruption from e.g., soft gaps and impurities will be removed by dissipation.
  In (b) and (c) temperatures of the blue and red curves equal $\omega_0/50$ and $\omega_0/150$, respectively.
  Assuming $\omega_0 \sim 3$meV, the observation of these two curves require a STM resolution of $0.06$meV and $0.02$meV, respectively, both realizable in real experiments~\cite{HanaguriRevSciIns18,MartaRevSciIns21}.
  }
  \label{fig:continuous_abs}
\end{figure*}

The above protocol to detect MZM with the conductance landscape however does not apply to the special case (see e.g., Ref.~\cite{JJinFeng-2016-PRL}) where the MZM state is the only discrete level within the superconducting gap.
Of this case, the dissipation effect in Majorana detection cannot be benchmarked by the splitting of a CdGM peak.
One can instead confirm the Majorana signal by investigating the dissipation-dependent universality class (i.e., the conductance scaling features) of the Majorana ZBCP.
Briefly, near the weak tunneling regime, dissipative tunneling into a MZM displays the scaling feature of the conductance $G_\text{MZM} \propto \epsilon^{2r - 1}$~\cite{DongPRL13}, where $r \equiv R e^2/h$ refers to the dimensionless dissipation.
The system energy $\epsilon$ can be either temperature or bias. We thus have two possible choices to observe the anticipated scaling feature: by changing either the system temperature $T$ [Fig.~\ref{fig:universal_features}(a) or tip-bias $V_\text{bias}$ [Fig.~\ref{fig:universal_features}(b)].
In these two cases, we anticipate to observe the scaling features in temperature ($T^{2r - 1}$) and bias ($V_\text{bias}^{2r - 1}$), respectively.
As the third possibility, we can also investigate the dependence of the ZBCP on the tip-sample distance [Fig.~\ref{fig:universal_features}(c)]. 
This option however 
requires to know the tip-sample tunneling amplitude as a function of the tip-sample distance. In making Fig.~\ref{fig:universal_features}(c), we simply assumed a Gaussian decaying (with width $\lambda$) of the tip-sample tunneling. In addition, scaling features emerge at relatively high energies. Their observation thus has a relatively milder requirement on the STM resolution.



Measurement of the scaling feature also helps confirming a MZM of generic cases (i.e., the cases with multiple CdGMs within the superconducting gap).
Indeed, with both pieces of evidence observed (i.e., the scaling feature of the ZBCP and the splitting of CdGM peaks), the existence of the MZM can be more strongly confirmed.
However, for generic cases, the observation of the scaling feature in bias requires a large-enough CdGM energy: otherwise the Majorana ZBCP scaling feature in bias is disguised by the CdGM contribution.

In previous paragraphs, we assume CdGMs that are distinguishable in energy.
For a vortex with CdGM energy level spacing smaller than the STM resolution, those CdGMs instead form into a continuous spectrum described by the density of states $\rho(\epsilon)$.
Of this case, dissipation also benefits the identification of the Majorana signal. Briefly, when the tunneling is dissipation free, the Majorana tunneling signal is disguised by the CdGMs around zero energy [Fig.\ref{fig:continuous_abs}(a)].
Here we have assumed a Gaussian-shape CdGM density of states~\cite{ShorePRL89} $\rho(\omega) = \rho(0) \exp(4 \omega^2/\omega^2_0)$, where $\omega_0 \approx 0.8 \Delta_\text{SC}$.
With dissipation included, in contrast, the CdGM tunnelings are strongly suppressed at low energies, thus manifesting the Majorana signal by a sharp ZBCP, especially under lower temperatures [Figs.~\ref{fig:continuous_abs}(b) and \ref{fig:continuous_abs}(c)].
Noticeably, to observe this feature, the distance between two ``shoulders'' of the curve, $\sim \omega_0\sqrt{r}$ should be much larger than the temperature $T$. This requirement is normally satisfied as $\omega_0 \gg T$. On the contrary, for the cases without MZM, dissipation always strongly suppresses the zero bias conductance and leads to a clean valley structure~\cite{DongPRL13,DonghaoPRL22}.

\begin{figure*}[t]
  \includegraphics[width=2\columnwidth]{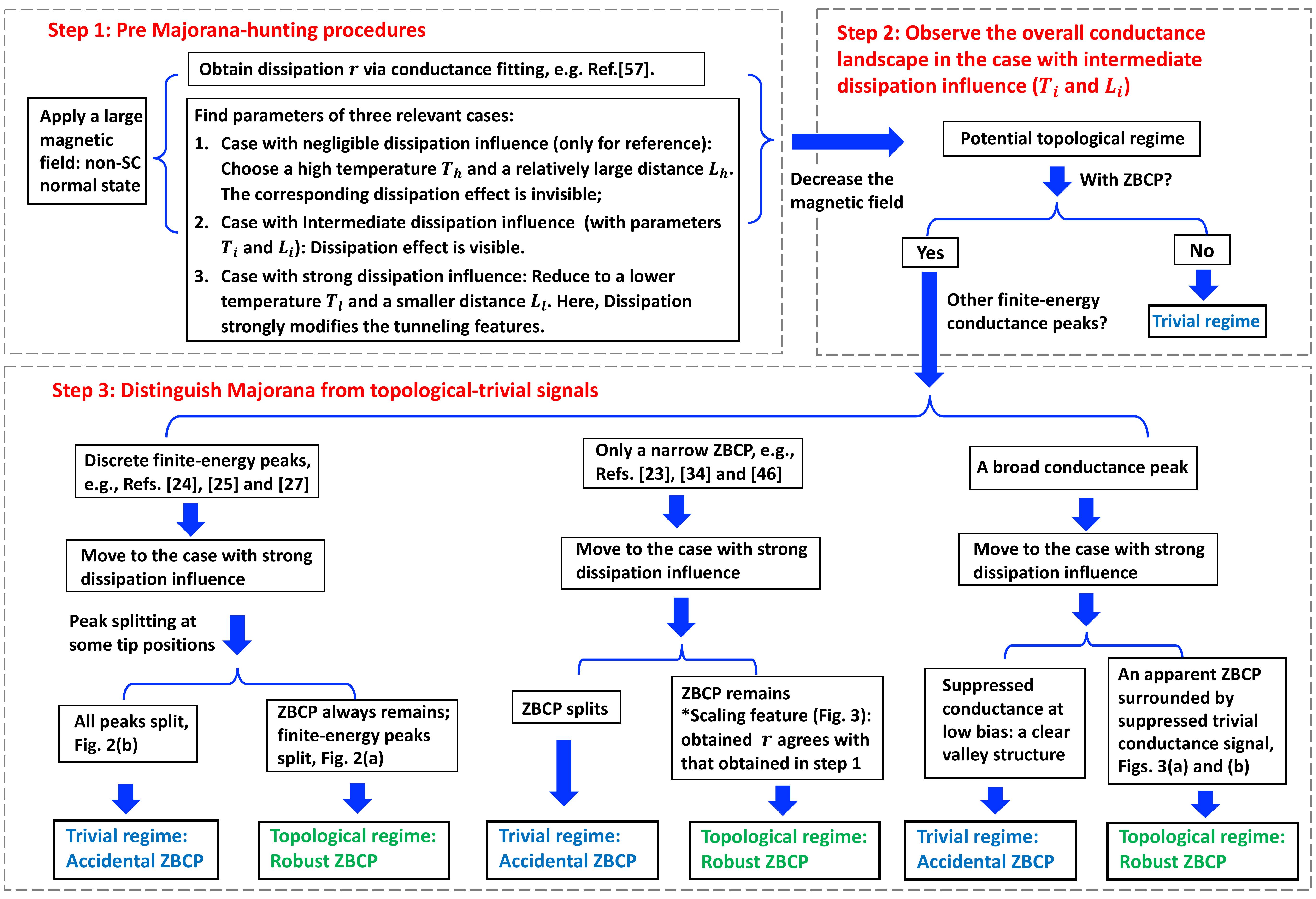}
  \caption{The protocol to detect Majorana in a dissipative vortex. We emphasize that the coincidence between dissipation $r$ obtained from scaling features (marked by the asterisk) and that from a direct measurement is a confirmative, yet not strictly necessary Majorana signature. Indeed, of the single-conductance-peak situation, the non-splitting zero-bias peak is by itself a strong piece of Majorana evidence.}
  \label{Protocol-Fig}
\end{figure*}

Based on the strong Majorana evidence discussed above, we provide a protocol to identify Majorana in a vortex, in Fig.~\ref{Protocol-Fig}.

\section{The system Hamiltonian and calculation details}

A vortex of the considered topological system (Fig.~\ref{fig:sample}) contains a MZM and multiple finite-energy CdGMs.
Tunneling (from the tip) into each of these localized states can be described by the Hamiltonian
\begin{equation}
\begin{aligned}
    H_\text{T} (\mathbf{r}) & = \sum_n\left[ t_{e\uparrow,n} (\mathbf{r}) \psi^\dagger_\uparrow (\mathbf{r}) a_n + t_{e\downarrow,n} (\mathbf{r}) \psi^\dagger_\downarrow  (\mathbf{r}) a_n \right.\\
    & \left. + t_{h\uparrow,n} (\mathbf{r}) \psi^\dagger_\uparrow(\mathbf{r}) a_n^\dagger + t_{h\downarrow,n} (\mathbf{r}) \psi^\dagger_\downarrow(\mathbf{r}) a_n^\dagger \right] e^{-i\varphi} + h.c.,
\end{aligned}
\label{eq:ht_abs}
\end{equation}
where $\mathbf{r}$ labels the (three dimension) tip position,
$\psi_{\sigma}$ refers to the tip fermion with spin $\sigma$, and $a_n$ refers to the vortex state ($n=0$ for the Majorana and otherwise for a CdGM).
The electron and hole sectors of a local state are labelled by $e$ and $h$, respectively.
In Eq.~\eqref{eq:ht_abs}, $\varphi$ is the phase
fluctuation operator (of the tip-sample junction) that couples to dissipative environmental modes. This phase is conjugate to the charge number of the tip-sample junction, with its dynamics given by the long-time correlation~\cite{DevoretPRL90}
\begin{equation}
\langle \exp[i\varphi (t)] \exp[-i\varphi (0)] \rangle \propto t^{-2r},
\label{eq:dissipation_dynamics}
\end{equation}
where $r = R e^2/h$ is the dimensionless dissipation amplitude.
The dynamics Eq.~\eqref{eq:dissipation_dynamics} indeed agrees with experimental data obtained in a Pb-superconducting island, where dissipation is introduced via the disordered layer~\cite{RoditchevPRL13,JeonKimKukNJP20}

Especially, for the tip-Majorana coupling,
parameters $|t_{e\uparrow,0}(\mathbf{r}) |^2 + |t_{e\downarrow,0}(\mathbf{r})|^2 = |t_{h\uparrow,0}(\mathbf{r})|^2 + |t_{h\downarrow,0}(\mathbf{r})|^2$, due to the equality of electron and hole weights of a Majorana~\cite{Read-2000-PRB,Kitaev2001}, for any position $\mathbf{r}$.
The tip-Majorana coupling then becomes
\begin{equation}
H_\text{T-MZM} (\mathbf{r}) = t_\text{MZM} (\mathbf{r}) \left( \psi_\text{tip}^\dagger e^{-i\varphi} - \psi_\text{tip} e^{i\varphi} \right) \gamma,
\label{eq:ht_mbs} 
\end{equation}
where $\gamma = \gamma^\dagger = (a_0 + a_0^\dagger)/\sqrt{2}$ is the vortex MZM.
Due to its polynomial long-time feature Eq.~\eqref{eq:dissipation_dynamics}, phase $\varphi$ can be combined with the field $\phi_\text{tip}$ introduced from the bosonization of the tip fermion $\psi_\text{tip} \sim \exp(i\phi_\text{tip})/\sqrt{2\pi a}$~\cite{KaneFisherPRB92}, where $a$ is the short-distance cutoff. By doing so, the tip-Majorana coupling instead becomes
\begin{equation}
H_\text{T-MZM}' = -it_\text{MZM} \sqrt{\frac{2}{\pi a}} \sin (\sqrt{1 + 2r} \Phi) \gamma,
\label{eq:bsg}
\end{equation}
with the combined phase $\Phi = (\phi_\text{tip} + \varphi)/\sqrt{1 + 2r}$ and the MZM-coupling strength $t_\text{MZM}$.
The operator of Eq.~\eqref{eq:bsg} has the scaling dimension $1/2 + r$, and is thus relevant as long as $r<1/2$.
In this work, ``relevance'' is defined in the renormalization group (RG) perspective. An tunneling is relevant (irrelevant), if its tunneling parameter (e.g., $t_\text{MZM}$) becomes increasingly important (unimportant) when decreasing the system energy (e.g., temperature or bias)~\cite{KaneFisherPRB92}.
The Majorana fermion $\gamma^2= 1$ and couples only to the tip. The tip-Majorana tunneling Eq.~\eqref{eq:bsg} is then equivalent to the Hamiltonian of a boundary Sine-Gordon model~\cite{GhoshalZamolodchikov94,FendleyPRB95}, which is exactly solvable with thermodynamic Betha Ansatz~\cite{FendleyPRB95} (see a brief outline in Appendix~\ref{app:tba}). 
Following Ref.~\cite{FendleyPRB95}, we work out the MZM conductance in Figs.~\ref{fig:universal_vs_non-universal}(a) and \ref{fig:universal_features} (for $r = 0.2$).
Especially, the conductance presents perfect scaling features as functions of temperature [Fig.~\ref{fig:universal_features}(a)] and bias [Fig.~\ref{fig:universal_features}(b)], respectively.
Following the results of the boundary Sine-Gordon~\cite{FendleyPRB95}, the MZM conductance always increases~\cite{DongPRL13}, when decreasing the temperature, the applied bias, or the tip-sample distance: as long as the tunneling model~\eqref{eq:bsg} remains valid.


In contrast to the Majorana situation, generically the tip-CdGM tunneling involves four independent tunneling parameters as shown in Eq.~\eqref{eq:ht_abs}. The tip-CdGM tunneling then can not reduce to the analytically solvable Hamiltonian Eq.~\eqref{eq:bsg}.
Now, we perturbatively investigate the tunneling into CdGMs with the Coulomb-gas RG method (see Appendix~\ref{app:coulomb-gas}).
Briefly, with RG equations, one obtains the effective tunneling parameters under different energies $\epsilon$. In real experiments, $\epsilon$ can be either bias $V_\text{bias}$ in the tip or the system temperature $T$.
At the beginning of the RG flow, all four tunneling parameters follow the same flow equation
\begin{equation}
\frac{d t_\alpha}{dl} = \left[ 1 - \left( \frac{1}{2} + r \right) \right] t_\alpha,
\label{eq:transmission_flows}
\end{equation}
for $\alpha = e\uparrow, h\uparrow, e\downarrow$ and $h\downarrow$, and all $n$ values alike.
The parameter $l = \exp(\epsilon_0/\epsilon) - 1$ labels the progress in the RG flow, where $\epsilon_0$ is the initial RG cutoff. It equals zero initially (where $\epsilon = \epsilon_0$), and increases with a decreasing RG cutoff $\epsilon$.
In real experiments, $\epsilon$ refers to the energy fluctuation of the system. Specifically, it can be considered as either the difference between local-state energy and the applied bias, or the system temperature, whichever is larger.
Following Eq.~\eqref{eq:transmission_flows}, when RG flow begins, all tunneling parameters share the scaling dimension of the Majorana-tunneling $r + 1/2$ of Eq.~\eqref{eq:bsg}.
They are all relevant if $r < 1/2$, and effectively increase when $\epsilon$ decreases.
Scaling dimensions of them however become different after the RG flow (see Appendix~\ref{app:coulomb-gas}).
More specifically, the system generically begins to prefer the tunneling process with the leading amplitude.
At the end of the RG flow, the system approaches the weak tunneling fixed point, where only the tunneling process with the largest initial value dominates at low-enough energies.
For instance, near the fixed point where $t_{e\uparrow}$ dominates (i.e., $t_{e\uparrow} \gg t_{e\downarrow},t_{h\uparrow},t_{h\downarrow}$), the flow of tunneling parameters follows (here we drop the label of $n$ for simplicity)
\begin{equation}
\begin{aligned}
&\frac{dt_{e\uparrow}}{d l} = t_{e\uparrow},\ \frac{dt_{h\uparrow}}{d l} =- (1 + 4 r) t_{h\uparrow},\\ &\frac{dt_{e\downarrow}}{d l} = 0\  t_{e\downarrow},\ \frac{dt_{h\downarrow}}{d l} = - 4 r t_{h\downarrow},
\end{aligned}
\label{eq:rg_final}
\end{equation}
where only the leading process $t_{e\uparrow}$ remains relevant, and approaches perfect transmission after the RG flow.
Notice that a finite current can only be produced via the successive application of $e$ and $h$-involved tunneling processes.
Of the $t_{e\uparrow}$-dominating situation, the flow of current is then determined by $t_{h\downarrow}$, i.e., $G/(e^2/h) \propto 2t_{e\uparrow}^2 t_{h\downarrow}^2/(t_{e\uparrow}^2 + t_{h\downarrow}^2) \sim 2 t_{h\downarrow}^2$ (see Appendix~\ref{app:coulomb-gas}), where $t_{e\uparrow}\gg t_{h\downarrow}\gg$ other tunnelings near the fixed point. This combined process of $G$ is irrelevant in the RG, with the scaling dimension $1 + 4r$.
This irrelevance of current tunneling, remarkably, is distinct from the relevance of tunneling before the flow Eq.~\eqref{eq:transmission_flows}.
More specifically, transmission is a relevant process initially, where the conductance increases with a decreasing RG cutoff $\epsilon$.
After the RG flow, however, transmission becomes irrelevant, where conductance instead decreases with $\epsilon$.
A critical energy is thus anticipated as the ``watershed'' that separates the regimes with relevant and irrelevant transmission. Experimentally, this critical energy corresponds to the temperature where peak splitting begins, i.e. Fig.~\ref{fig:universal_vs_non-universal}.

Finally, when the island contains a continuous CdGM spectrum, the dissipative conductance can be obtained with the so-called $P(E)$ method~\cite{BrunPRL12,RoditchevPRL13,AstNatCom16}
\begin{equation}
\begin{aligned}
    & \Gamma_{\text{tip}\to\text{sample}}(V_\text{bias}) =  \frac{1}{R_\text{T}} \int d E \int dE' P(E) \\
    \times & \rho (E) \rho_\text{tip} (eV_\text{bias} + E') f (eV_\text{bias} + E') [1 - f (E)],
\end{aligned}
\label{eq:pe}
\end{equation}
with which the current equals $I(V_\text{bias}) = \Gamma_{\text{tip}\to\text{sample}}(V_\text{bias}) - \Gamma_{\text{sample}\to\text{tip}}(V_\text{bias})$. Here $R_\text{T}$ is the tunneling resistance (indicating the tunneling barrier strength), $\rho$ is the density of state of the sample at the tip location, $\rho_\text{tip}$ is the tip density of state, $f(E)$ is a fermionic distribution, and finally, $P(E)$ incorporates the chance to dissipate electron energy $E$ via the dissipative environment~\cite{Ingold1992SUS}.
For a dissipation-free system, the Majorana is fully disguised by the surrounding CdGMs, and simply increases $\rho$ of the island [Fig.~\ref{fig:continuous_abs}(a)]. It is hard to tell whether the conductance ``peak'' has any Majorana-origination.
With dissipation involved, the interrupting CdGM signal is strongly suppressed, leaving the Majorana signal standing out [Fig.~\ref{fig:continuous_abs}(b) and (c)].



\section{Discussion and summary}

In this paper, we have studied the transport features of a dissipative tunneling into a topological vortex.
Instead of the $2e^2/h$ conductance quanta, we propose three dissipation-unique phenomena that can help the identification of a Majorana.
\textbf{Firstly}, the CdGM peak can split when the tip approaches the sample (see Fig.\ref{fig:universal_vs_non-universal}).
This non-universal feature of the CdGM peak, if coexisting with a non-splitting ZBCP, is a strong Majorana signal.
The observation of this phenomenon requires the existence of finite-energy CdGMs.
Observation at different tip positions is also required, to avoid accidental equality of electron and hole weights (of CdGMs).
In addition, the STM resolution is preferred to be smaller than the temperature at which the peak splitting begins. Otherwise the peak-splitting can only be indirectly detected by measuring the CdGM peak half-width: which is expected to become larger if peak splitting occurs. For an sample with transmission $0.25$ at temperature $T = 10$K, the peak splitting is expected to occur around $1$K, larger than the latest STM resolution~\cite{HanaguriRevSciIns18,MartaRevSciIns21}.
\textbf{Secondly}, one can further confirm the Majorana existence by measuring the universal feature, i.e., the dependence of the ZBCP as functions of temperature [zero-bias, Fig.\ref{fig:universal_features}(a)] or bias [temperature smaller than bias, Fig.\ref{fig:universal_features}(b)].
Measuring the power-law scaling feature requires the STM resolution to be smaller than temperature/bias of the universal regime.
In addition, if one measures the scaling feature with bias, the applied bias is required to be smaller than the lowest CdGM energy: otherwise the scaling feature is interrupted by the CdGM conductance.
\textbf{Thirdly}, with interrupting CdGM signals suppressed by dissipation, the Majorana signal becomes manifest and experimentally accessible even when the CdGM energy difference is smaller than the STM resolution (Fig.~\ref{fig:continuous_abs}).
The STM resolution is only required to be smaller than the energy difference between ``shoulders'' of the continuous spectrum, which is normally at the order of the superconducting gap. 
Remarkably, all proposed phenomena do not require knowing the exact unit of conductance. A complicated calibration (of the measured current) is then unnecessary, lowering potential experimental difficulty.

Before closure, we briefly discuss the relevant experimental parameters, including the dissipation amplitude $r$ and the charging energy of the sample.
To begin with, we choose the dissipation $r \sim 0.2$ since such a dissipation amplitude has been proven as enough to suppress false positive signals (i.e., conductance peaks due to Andreev bound states) in a hybrid nanowire system~\cite{ShanZhangPRL22,ZhichuanPRB22}, at an experimentally accessible temperature $\sim 200$mK.
This dissipation amplitude has indeed been realized experimentally in a Pb superconducting island~\cite{BrunPRL12,RoditchevPRL13,JeonKimKukNJP20}.
For instance, Ref.~\cite{JeonKimKukNJP20} reports a dissipation $r \sim 0.26$ of a Pb-superconducting island due to a disordered Pb wetting layer. Similar dissipation amplitude is anticipated as experimentally realizable in a vortex system, by a disordered layer (Fig.~\ref{fig:sample}).
As the second requirement, the charging energy of the island should be smaller than the temperature required to see the peak splitting: otherwise the observation of the peak splitting can be interrupted by the Coulomb blockade effect.
In real experiments, a small charging energy
\begin{equation}
    E_c = \frac{e^2}{4\pi \varepsilon_\text{sample} \varepsilon_0 R_\text{sample}}
\end{equation}
requires a large sample radius $R_\text{sample}$ and large (relative) dielectric constant $\varepsilon_\text{sample}$ (here $\varepsilon_0$ is the vacuum dielectric constant).
For instance, a 2D sample with radius $R_\text{sample} \sim 2\mu$m is required to have a relative dielectric constant around $\varepsilon_\text{sample} \sim 50 $, to produce a small enough charging energy $E_c \sim 10\mu$eV $~ \sim 100$mK.
Fortunately, Majorana hunting experiments in iron-based vortexes, e.g., Refs.~\cite{FDongLai-2018-PRX,Hanaguri-2019-NatM,ZPeng-2019-NatPhys,KLingYuan-2019-NatPhys,WHaiHu-2019-arXiv,DHong-2020-NatCom,FDongLai-2021-PRL,KLingyuan-2021-NatCom,LWenyao-2021-arXiv} do not report apparent interruption from Coulomb blockade, thus supporting the applicability of our theory.
As a summary, it is experimentally feasible to fabricate a sample that is both free from Coulomb blockade, and also has enough dissipation to generate our predicted phenomena.


\begin{appendix}

\section{Coulomb-gas RG equations}
\label{app:coulomb-gas}

Briefly, the renormalization group (RG) method investigates the low-energy system features (e.g., the conductance) by figuring out the change of system parameters (the so call ``RG flow'') when decreasing the system energy (temperature, bias, or the inverse of some dephasing time).
In general, a RG method deals with the system action with the form~\cite{AltlandBook}
\begin{equation}
    S_\text{system} [\phi,\Lambda,\left\{ \gamma_\alpha \right\}] = \sum_{\alpha = 1}^N \gamma_\alpha \mathcal{O}_\alpha [\phi],
\end{equation}
where $\phi$ is some field, and $\gamma_\alpha$ is the prefactor of the operator $\mathcal{O}_\alpha$ labelled by $\alpha$.
The system has an initial cutoff $\Lambda$ in energy at the beginning of the RG flow.
When this cutoff decreases $\Lambda\to \Lambda - \delta \Lambda$, particle states with energies between the initial and reduced cutoffs should be removed, after effectively incorporating the interaction between these states and the states with energies below the new cutoff.
These interactions then effectively modify the prefactors $\gamma_\alpha$, leading to an effectively new action $S_\text{system} [\phi, \Lambda- \delta \Lambda,\left\{ \gamma_\alpha' \right\}]$.
When $\delta \Lambda$ is infinitesimal, the change of parameters $\gamma_\alpha$ follows the RG flow equations.
This basic idea applies to most RG methods, including the Coulomb gas RG.

More specifically, for the Coulomb gas RG method, one deals with the partition function
\begin{equation}
\begin{aligned}
& \mathcal{Z} (\tau_c, \left\{\gamma_\alpha \mathcal{O}_\alpha\right\})  = \sum_n\int_0^{\beta} d\tau_1 \cdots \int_0^{\tau_{i+2}-\tau_c} d\tau_{i+1} \cdots  \\
& \times \Big\langle  \cdots H_{\text{T}}(\tau_{i}, \left\{\gamma_\alpha \mathcal{O}_\alpha\right\}) H_{\text{T}}(\tau_{i+1},\left\{\gamma_\alpha \mathcal{O}_\alpha\right\}) \cdots \Big\rangle_0
\end{aligned}
\label{seq:partition_function}
\end{equation}
that is required to be invariant during the RG flow.
In Eq.~\eqref{seq:partition_function}, the integral of ``time'' is between zero and the temperature inverse $\beta = 1/T$.
As a consequence, strictly Coulomb gas RG equations describe the flow of equilibrium systems by decreasing the temperature.
However, if within the weak-tunneling regime, temperature and bias are exchangeable as both can be considered as the energy cutoff.
Indeed, difference between temperature and bias emerges only in the crossover regime~\cite{MebrahtuNaturePhy}, i.e., away from both the strong and weak coupling fixed points.

The partition function Eq.~\eqref{seq:partition_function} contains ``kinks'' at moments $\tau_1, \tau_2 ... \tau_i ... $ when the system state changes.
In the Coulomb gas RG method, a cutoff in time $\tau_c$ is introduced.
The Coulomb gas RG contains the steps below.
To begin with, before each RG step, the ``time difference'' $|\tau_i - \tau_{i+1} | > \tau_c$ between any two kinks is assumed to be larger than the cutoff $\tau_c$, forbidding the occurrence of two kinks within $\tau_c$.
When the RG step begins, this cutoff increases to $\tau_c \to \tau_c + \delta \tau_c$, where $\delta \tau_c > 0$ refers to the changing of the cutoff.
Notice that $1/\tau_c$ can be interpreted as the cutoff in energy. The increasing of $\tau_c$ then indicates the decreasing of the energy cutoff.
Due to the increasing of cutoff in time, kinks now have the chance to occur within the new cutoff, i.e., $\tau_c < |\tau_i - \tau_{i+1}| < \tau_c + \delta \tau_c$.
Contribution of these kinks should then be integrated out within this RG step.
The contribution of the integrated pairs to the partition function then leads to the modification of the partition function $\mathcal{Z} (\tau_c, \left\{\gamma_\alpha \mathcal{O}_\alpha\right\}) \to \mathcal{Z} (\tau_c + \delta \tau_c, \left\{\gamma'_\alpha \mathcal{O}_\alpha\right\})$, with a set of modified parameters $\gamma'_\alpha$.
Finally, to guarantee the invariance of partition function, we rescale the cutoff back to $\tau_c$. When $\delta \tau_c \ll \tau_c$ is infinitesimal, the modification of parameters can be written in terms of differential equations known as RG equations.

Now we go back to our specific system i.e., the dissipative tunneling into the vortex.
Experimentally, the realization of this system is distinct from that of a hybrid nanowire.
However, the tunneling Hamiltonian Eq.~\eqref{eq:ht_abs} is exactly the same as the dissipative tunneling into an Andreev bound state of a superconducting-proximitized nanowire.
The RG equations of the hybrid nanowire system~\cite{DonghaoPRL22} are thus also capable to describe the flow of parameters of the vortex system under consideration.
As the only difference, Ref.~\cite{DonghaoPRL22} considers only the zero-bias conductance peak, where the Andreev bound state energy serves as another possible energy cutoff.
In our case, by contrast, non-equlibrum tunneling into a CdGM with finite energy $\epsilon_\text{CdGM}$ is considered.
This problem can however be solved by treating the bias that equals the CdGM energy $V_\text{bias} = \epsilon_\text{CdGM}$ as the effective zero bias.
We emphasize that the Majorana identification protocol in our work strongly relies on the vortex-specified feature where finite-energy CdGMs and a MZM coexist. This feature is novel as it does not appear in hybrid nanowire systems~\cite{DonghaoPRL22}.
With this treatment, tunneling parameters follow the RG equations

\begin{equation}
\begin{aligned}
\frac{dt_{e\uparrow}}{dl} & = \left[ 1 - \frac{(K_1 + 1)^2 + g (K_2 + 1)^2}{4g} \right] t_{e\uparrow} , \\ \frac{dt_{h\uparrow}}{dl}  &= \left[ 1 - \frac{(K_1 - 1)^2 + g (K_2 - 1)^2}{4g} \right] t_{h\uparrow}, \\
\frac{dt_{e\downarrow}}{dl} & = \left[ 1 - \frac{(K_1 + 1)^2 + g (K_2 - 1)^2}{4g} \right] t_{e\downarrow} , \\ \frac{dt_{h\downarrow}}{dl} & = \left[ 1 - \frac{(K_1 - 1)^2 + g (K_2 + 1)^2}{4g} \right] t_{h\downarrow},
\end{aligned}
\label{seq:transmission_flows}
\end{equation}
where $g = 1/(1 + 4r)$, and $l \propto \ln \epsilon$ labels the RG process in terms of the energy $\epsilon$.
In experiments, $\epsilon$ can be considered as temperature or bias.
Initially, $l=0$, where both parameters $K_1(l=0) = K_2 (l = 0) = 0$ equal zero. At this point, all tunneling parameters of Eq.~\eqref{seq:transmission_flows} follow the same flow equation, i.e., $dt_\alpha/d l = [1 - (r + 1/2)] t_\alpha$ for $\alpha = e\uparrow, h\uparrow, e\downarrow$ and $h\downarrow$.
The values of $K_1$ and $K_2$ begin to change when $l$ increases, following
\begin{equation}
\begin{aligned}
\frac{d K_1}{dl} & = -2 \tau_c^2 \left[ ( |t_{e\uparrow}|^2 - |t_{h\uparrow}|^2 + |t_{e\downarrow}|^2 - |t_{h\downarrow}|^2 ) \right. \\
& \left. + ( |t_{e\uparrow}|^2 + |t_{h\uparrow}|^2 + |t_{e\downarrow}|^2 + |t_{h\downarrow}|^2 ) K_1 \right], \\
\frac{d K_2}{dl} & = -2 \tau_c^2 \left[ ( |t_{e\uparrow}|^2 - |t_{h\uparrow}|^2 - |t_{e\downarrow}|^2 + |t_{h\downarrow}|^2 ) \right. \\
& \left. + ( |t_{e\uparrow}|^2 + |t_{h\uparrow}|^2 + |t_{e\downarrow}|^2 + |t_{h\downarrow}|^2 ) K_2 \right].
\end{aligned}
\label{seq:fugacity_rg}
\end{equation}
Noticeably, Eq.~\eqref{seq:fugacity_rg} is asymmetric with respect to tunneling parameters, through which the final values of $K_1$ and $K_2$ are determined by the leading process.
For instance, with $t_{e\uparrow}$ dominant, $K_1 = K_2 = -1$ after the RG flow, leading to a most strongly reduced scaling dimension of $t_{e\uparrow}$ [Eq.~\eqref{eq:rg_final}].

By solving flow equations~\eqref{seq:transmission_flows} and \eqref{seq:fugacity_rg}, and the conductance expression
\begin{equation}
    G \propto \frac{e^2}{h} \tau_c^2 \left( \frac{2t_{e\uparrow}^2 t_{h\downarrow}^2}{t_{e\uparrow}^2 + t_{h\downarrow}^2} + \frac{2t_{e\downarrow}^2 t_{h\uparrow}^2}{t_{e\downarrow}^2 + t_{h\uparrow}^2}\right),
\end{equation}
we arrive at the CdGM conductance peak contributions in Fig.~\ref{fig:universal_vs_non-universal} (see the contained Mathematica Notebook). 
For a generic CdGM, one of the four parameters dominates after the flow. For instance, if $t_{e\uparrow}$ dominates, the conductance $G \propto 2t_{e\uparrow}^2 t_{h\downarrow}^2/(t_{e\uparrow}^2 + t_{h\downarrow}^2)$.

\
\
\
\

\section{Thermaldynamic Bethe Ansatz}
\label{app:tba}

As has been discussed, since $\gamma^2 = 1$, Eq.~\eqref{eq:bsg} agrees with that of a boundary Sine Gordon model. It has been shown that~\cite{FendleyPRB95} such a system is perfectly integrable (and thus exactly solvable), and contains quasiparticles including a kink, an anti-kink, and may have multiple breathers.
To figure out the number of breathers, we notice that our Hamiltonian $\sim \sin (\sqrt{1 + 2r} \Phi)$ has the scaling dimension $1/2 + r$. Since $1 < 1/(1/2 + r) < 2$ if $r < 1/2$, following Ref.~\cite{FendleyPRB95}, the system we consider has zero breather.

To proceed, one needs to solve the dispersion function of quasiparticles (kink and anti-kink), as functions of the rapidity $\theta$. More specifically, the dispersion function can be figured out from the integral equation~\cite{FendleyPRB95}
\begin{equation}
\begin{aligned}
    & \epsilon_\beta (\theta,V_\text{bias}) = \frac{M e^\theta}{T} - \sum_{\beta'} \int_{-\infty}^\infty d\theta \mathbf{\Phi}_{\beta,\beta'} (\theta - \theta')\\
    \times & \ln \left[ 1 + e^{\beta' V_\text{bias} /2 T - \epsilon_{\beta'} (\theta, V_\text{bias}/T)} \right],
    \label{seq:dispersion_integral}
\end{aligned}
\end{equation}
where $\beta,\beta'$ equal 1 and $-1$ for kink and anti-kink, respectively.
The function $\mathbf{\Phi}_{\beta,\beta'}$ is related to the quasiparticle scattering matrix in the bulk
\begin{equation}
    \mathbf{\Phi}_{\beta,\beta'} (\theta) = -\frac{1}{2\pi\cosh\theta}.
\end{equation}
The first term of Eq.~\eqref{seq:dispersion_integral}, i.e., $M \exp(\theta)/T$ influences the interaction-free contribution. The choice of $M$ is a common prefactor that will not influence the final conductance.

With the dispersion function obtained, the current equals
\begin{widetext}
\begin{equation}
\begin{aligned}
    I(T_\text{ref},V_\text{bias},T) = \frac{e T n(r)}{2 h} \int_{-\infty}^\infty d\theta \frac{1}{\cosh^2\left\{ (r - 1/2) \left[ \theta - \ln (T_\text{ref} /T) \right] \right\}} \ln\left[ \frac{1 + e^{V_\text{bias}/2T - \epsilon (\theta + \ln (M/2T), V_\text{bias}) }}{1 + e^{-V_\text{bias}/2T - \epsilon (\theta + \ln (M/2T), V_\text{bias}) }} \right],
    \label{seq:current}
\end{aligned}
\end{equation}
\end{widetext}

where $T_\text{ref}$ is the effective ``reflection'' amplitude.
When $T_\text{ref} /T \ll 1$, reflection is weak, and conductance approaches the quantized value $2e^2/h$. In the opposite limit $T_\text{ref} /T \gg 1$, reflection is strong, and conductance approaches zero.
A large $T_\text{ref}$ indicates a small tip-sample tunneling, or a large tip-sample distance.
In Eq.~\eqref{seq:current}, the function $n(r)$ is a normalization factor.
Theoretically, it equals $2r + 1$. In real numerics, its value might deviates from $2r + 1$ by e.g., one to two percents. Due to the smallness of conductance in the weak tunneling limit, one needs to figure out the precise value of the normalization factor.

With Eq.~\eqref{seq:current}, we obtain the Majorana conductance peaks in Figs.~\ref{fig:universal_vs_non-universal} and \ref{fig:universal_features} (see C coding provided).

\section{Evaluation of the peak splitting energy}
\label{app:spliting_energy}

As an important question, experimentalists would wonder what is the approximate temperature or tip-sample distance when peak splitting occurs.

Without loss of generality, we consider the situation where $t_{e\uparrow}$ dominates after the RG flow.
Of this case, peak splitting occurs when the conjugate process $t_{h\downarrow}$ has become marginal in the RG prospective.
For simplicity, assuming equal $K_1$ and $K_2$, the RG-marginal of $t_{h\downarrow}$ requires $K_1 = K_2 = -(1 - 2r)/(1 + 2 r)$.
Following Eq.~\eqref{seq:fugacity_rg}
and noticing that $K_1(0) = K_2(0) = 0$ initially, the changing of $K_1$ and $K_2$ during the RG flow is triggered by the first term of Eq.~\eqref{seq:fugacity_rg}.
This term is small under a weak particle-hole asymmetry $ |t_{e\uparrow}|^2 + |t_{e\downarrow}|^2 \approx |t_{h\uparrow}|^2 + |t_{h\downarrow}|^2 $.
Of this situation, the flows of $K_1$ and $K_2$ are rather slow, and peak splitting can only occur at a significantly small temperature.

On the contrary, in the opposite limit where $|t_{e\uparrow}|\gg |t_{e\downarrow}| ,|t_{h\uparrow}| , |t_{h\downarrow}|$, the flow of $K_1$ and $K_2$ is basically determined by $t_{e\uparrow}$.
One can approximately predict the peak splitting by measuring the tip-sample transmission $\Gamma_\text{bm}$ at a temperature $T_\text{bm}$ that is high enough to avoid suppression from dissipation.
With this measurement, we ask what is the requirement to observe the peak splitting at the experimental temperature $T_\text{exp}$.
Here we further approximately treat $|t_{e\uparrow}|$ as a constant.
With these simplifications, at the experimental temperature $T_\text{exp}$, one anticipates the peak splitting, if the initial transmission $\Gamma_\text{bm}$ satisfies
\begin{equation}
    \Gamma_\text{bm} > \frac{\ln\frac{1+2r}{4r}}{\ln \frac{T_\text{bm}}{T_\text{exp}}}.
\end{equation}
From this equation, one sees that in the extremely asymmetric limit, peak splitting requires a large transmission (small tip-sample distance), or a small temperature $T_\text{exp}$.
Meanwhile, since $\ln [(1+2r)/(4r)]$ decreases monotonously when $r$ increases, the peak splitting of a sample with a larger $r$ occurs at a relatively larger temperature --- since peak splitting at smaller values of $K_1$ and $K_2$.
For a more intuitive understanding, if $r = 0.2$, $\Gamma_\text{bm} = 0.2$ when $T_\text{bm} = 5$K, then peak splitting is anticipated to occur around $T_\text{exp} \approx 500$mK.

\end{appendix}

\begin{acknowledgements}
The authors thank Hao Zhang and Zhan Cao for valuable discussions. This work was supported by China MOST’s Innovation Program for Quantum Science and Technology (Grant No.~2021ZD0302400, No.~2021ZD0302500, No.~2021ZD0302700), the National Natural Science Foundation of China (Grants No.~11974198, No.~12074133, NO.~11920101005), Tsinghua University Initiative Scientific Research Program, CAS Project for Young Scientists in Basic Research (YSBR-003), and the Strategic Priority Research Program of CAS (Grant No.~XDB28000000). 
\end{acknowledgements}

\bibliography{reference.bib}

\end{document}